# Disturbance Observer Application for the Compensation of the Phase Drift of the LANSCE DTL LINAC Solid State Power Amplifier*

Sungil Kwon[†], M. S. Barrueta, L. Castellano, J. M. Lyles, M. Prokop, P. Van Rooy, P. Torrez
Los Alamos National Laboratory, Los Alamos, NM, USA

*Abstract*

The front end of Los Alamos Neutron Science Center (LANSCE) linear accelerator uses four 201.25-MHz Drift-Tube Linacs (DTLs) to accelerate the H$^+$ and H$^-$ beams to 100 MeV. Three of the 201.25-MHz DTLs are powered by diacrodes and the first DTL is powered by a tetrode. A 20-kW solid-state power amplifier (SSPA) is used to provide ~15 kW drive power to the tetrode. The SSPA is water-cooled and consists of 24 push-pull LDMOS transistors operating at 45% of their power saturation capability, providing ample power headroom and excellent linearity. However, the phase of the SSPA is perturbed at +/-20 degrees over a few ten minutes partially caused by the temperature dependent phase variation of the air-cooled SSPA driver circulator. This phase variation consumes most of the phase control margin of the cavity field feedback controller. In order to mitigate the effect of the SSPA's phase variation on the cavity field, a disturbance observer controller (DOBC) has been designed and implemented on the cavity field control FPGA, which functions in parallel with the cavity field feedback controller. In this paper, the DOBC design and its function as well as its short- and long-term performance are addressed.

## 1. INTRODUCTION

The capabilities of the Los Alamos Neutron Science Center (LANSCE) experimental facilities include: 1) the Lujan Center, which requires short, high-intensity proton bunches in order to create short bursts of moderated neutrons with energies in the meV to keV range; 2) the Proton Radiography (pRad) Facility, which provides time-lapse images of dynamic phenomena in bulk material (for example, shock wave propagation) via 50-nsec-wide proton bursts, repeated at time intervals as short as 358 nsec with programmable burst repetition rates; 3) the Weapons Neutron Research (WNR) Facility, which provides unmoderated neutrons with energies in the keV to MeV range; 4) the Isotope Production Facility (IPF), which uses the 100-MeV H$^+$ beam to make medical radioisotopes; and 5) the Ultra Cold Neutron (UCN) Facility, which creates neutrons with energies below the $\mu eV$ energy range for basic physics research [1]. For all beam species, the beam pulse length is required to be adjusted. The default beam pulse length is 625 usec and it can be increased to 785 usec.

The ability of the digital low-level RF (DLLRF) control system to accommodate various beam loading conditions is crucial for successful LANSCE operations in which a wide variety of beam types of various levels of beam loading are present in the accelerator's RF cavities. To provide the stable cavity field before the beam loading and to minimize the perturbation of the cavity field caused by beam loading, the LANSCE DLLRF control system implements both a proportional-integral (PI) feedback controller (FBC) and feedforward controller capabilities.

For a small peak current beam loading, the PI FBC is sufficient to compensate for the beam loading in the cavity field. However, for high peak current beam loading, the simple PI FBC is not sufficient and a feedforward controller is crucial to the beam loading compensation capability. Furthermore, in order to keep the stability of the closed loop system against the external disturbance inputs, system uncertainties, the PI FBC should provide the gain and phase margin sufficiently.

At the LANSCE DTL tank1, a SSPA provides 15 kW RF power to the final stage high amplifier Thales TH781 tetrode. While the SSPA guarantees ample power margin and excellent linearity, it was observed that at the RF turn-on transient of a few ten minutes, its phase is perturbed at 40 degrees. This transient phase perturbation consumes most of the phase control margin of the PI FBC and increases the possibility of the RF trip. In this paper, a method to mitigate the effect of the phase variation on the cavity field stability is addressed. For this purpose, the phase variation of the SSPA is treated as input disturbance and a DOBC is designed, implemented on the DLLRF system to provide enough phase margin to the PI FBC.

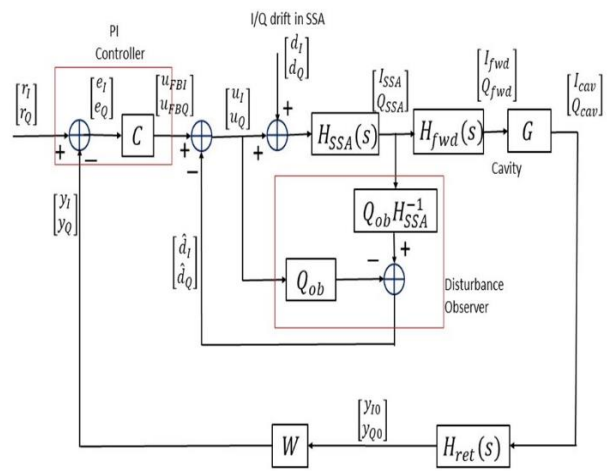

**Figure 1.** High-level functional diagram of the LANSCE digital low-level RF control system.

___________________
* Work supported by U.S. Dept. of Energy
† email address: skwon@lanl.gov.

## 2. BASEBAND UNCERTAINTY MODEL OF A RF SYSTEM

The baseband RF cavity in the In-phase(I) and Quadrature(Q) coordinate is modelled as a two-input-two-output (TITO) uncertain system caused by the unknown detuning frequency $\Delta\omega$ [2, 3]. The high power RF amplifier at an operating point can be modelled with a gain and two-by-two phase rotation matrix. The cavity RF feedback loop also can be modelled with a gain and two-by-two phase rotation matrix as well. As a result, overall LLRF control system can be described as a perturbed system $G_p(s)$ with the multiplicative uncertainty $\Delta(s, \Delta\omega)$ and shaping filter $M(s)$.

$$G_p(s) = G_n(s)(I + \Delta(s, \Delta\omega)M(s)). \quad (1)$$

In (1),

$$G_n(s) = \frac{h}{\tau_p s + 1}\begin{bmatrix} cos(\theta) & -sin(\theta) \\ sin(\theta) & cos(\theta) \end{bmatrix}, \quad (2)$$

$$\Delta(s, \Delta\omega) = \frac{1}{D(s, \Delta\omega)}\begin{bmatrix} -\left(\frac{\Delta\omega}{\omega_{3dB}}\right)^2 & -\frac{\Delta\omega}{\omega_{3dB}}(\tau_p s + 1) \\ \frac{\Delta\omega}{\omega_{3dB}}(\tau_p s + 1) & -\left(\frac{\Delta\omega}{\omega_{3dB}}\right)^2 \end{bmatrix},$$

(3)

$$D(s, \Delta\omega) = (\tau_p s + 1)^2 + \left(\frac{\Delta\omega}{\omega_{3dB}}\right)^2,$$

$$\|\Delta(j\omega, \Delta\omega)\|_\infty < 1,$$
$$\|\Delta(j\omega, \Delta\omega)\|_\infty < \|M(j\omega)\|_\infty,$$

and $\tau_p$ is the time constant of the cavity, $h$ is the steady state loop gain, $\theta$ is the phase rotation of the whole loop.

The characteristics of the perturbed system $G_p(s)$ is represented well in frequency domain. Bode plots show the magnitude response and the phase response of the system as well as the quantities of magnitude and phase of the crosstalk between the I and Q channels. Figure 2 shows the bode plots of the off-diagonal terms of the perturbed system $G_p(s)$, assuming that the phase rotation in the whole loop is zero ($\theta = 0$). Figure 2a is the frequency response from the Q channel input $u_Q$ to the I channel output $y_I$. Figure 2b is the frequency response from the I channel input $u_I$ to the Q channel output $y_Q$. They show the crosstalks between I channel and Q channel of $G_p(s)$ in both magnitude response and phase, which is caused by the detuning frequency $\Delta\omega$. The figures show that when the inputs are of broadband and the frequency spectra of them are swing from high frequency to low frequency, which occurs at the transient period of the pulsed inputs, the variations of the outputs of the amplitude and the phase of are ample. An approach to minimize the effects of the off-diagonal term outputs on the overall outputs is to apply high gain feedback controllers to the diagonal-terms so that the output portions of the diagonal-terms become dominant [4]. This approach works well when the system is a linear system and there is no limitation on the controller output magnitude. However, in our case, where the high peak reflected power of the cavity causes the RF trip, the application of the high gain PI feedback controller is not appropriate. Instead, an alternative approach is pursued. In LANSCE DLLRF system, the post-compensator that decouples the crosstalk, decomposing the TITO system into two Single-Input Single-Output(SISO) systems is designed and implemented [2, 5].

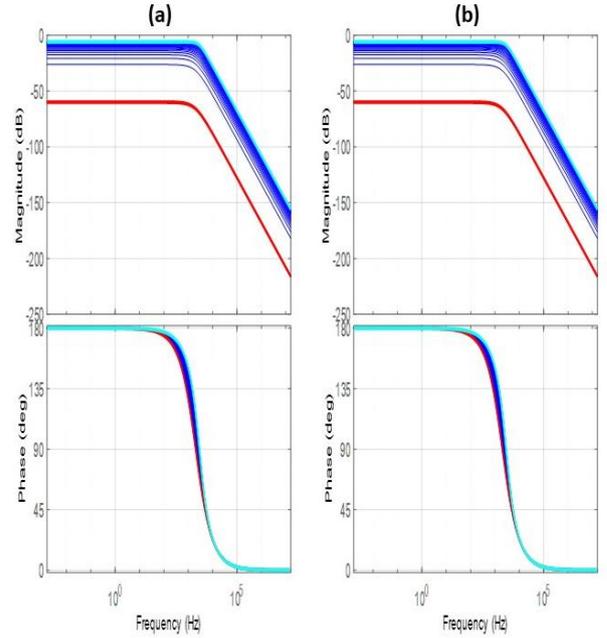

**Figure 2.** Bode Plots of the perturbed system: (a) $u_Q$ to $y_I$; (b) $u_Q$ to $y_I$. The 3 dB bandwidth of the cavity is 2013 Hz. The detuning frequency range for investigation is from 20 Hz to 2013 Hz. Red line: $\Delta f = 20$ Hz. Cyan line: $\Delta f = 2013$ Hz.

## 3. HIGH POWER RF AMPLIFIERS

The Thales TH781 tetrode operates in a cavity amplifier circuit provided by the same company. It is similar to the penultimate stage in the three previously upgraded DTL RF sources but with a larger 6-1/8 inch output and 1-5/8 inch input connections. Tube lifetime averages 47K hours for the TH781 at LANSCE, considered excellent for power grid tubes. In this application, we expect shorter filament emission life with higher peak cathode current, but with acceptable operation having the matched load of a circulator [6].

A 20 kW SSA from R&K Company provides ~15 kW of power to the tetrode final PA. This compact amplifier is water cooled, and consists of eight 3 kW pallets, each combining four MRF1K50H LDMOS transistors from NXP. The pallet outputs are combined in a 8-way radial coaxial combiner. The 24 push-pull LDMOS transistors operate at 45% of their saturated capability, so there is ample headroom and good linearity. The SSA power gain is 46 dB, and it is driven from a 500 milliwatt preamplifier in the LLRF rack [6].

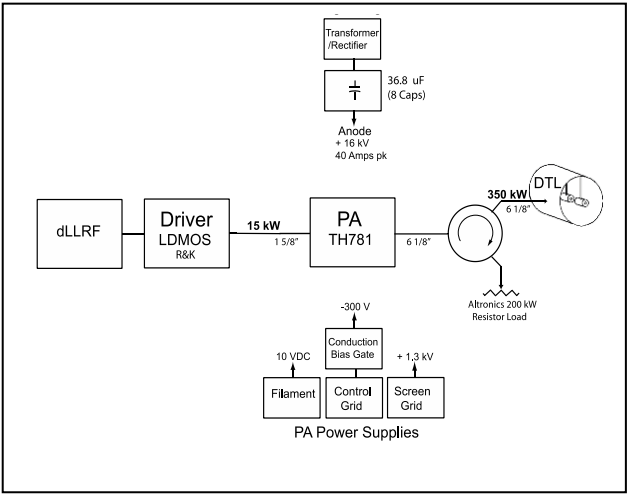

**Figure 3**. Diagram of New RF Amplifier System of DTL Tank1 at LANSCE LINAC

A slow degradation of stability in the LLRF PI controller was discovered during the first few tens minutes of RF system warm up. It would lead to severe overshoot when the PI loops locked at the start of flat top in each RF pulse. This overshoot delayed beam start time and caused reflected power faults. Using an independent pulsed phase measurement technique, we found that a slow phase change inside the SSA gradually destabilized the LLRF loop. It shifted more than 20 degrees of RF phase, to the edge of the phase margin of the PI controller [6]. In order to solve this problem, we implemented a DOBC on the DLLRF DSP, essentially a phase controller to normalize and hold the plant (amplifiers) phase within limits.

Figure 4 show phase variation of the SSPA. Data was captured at the middle of the RF 1000 usec RF pulse. Figure 4a shows 52 hour long phase drift. Figure 4b shows the phase variation monitored after the RF was recovered from a few hour turn off (zoom of box1 area in figure 4a). In this case, SSPA was fully cool down and for 40 mins of RF turn-on transient, 39 degrees of phase variation was observed. Figure 4c shows the phase variation monitored after the RF was recovered from the short term (<20 mins) trip (zoom of box2 area in figure 4b). In both cases, DOBC increased the phase margin of the PI FBC and the cavity field phase (green lines) remained stable.

## 4. DISTURBANCE OBSERVER CONTROLLER (DOBC)

The input-output representation of the SSPA system in the I/Q coordinate can be expressed as

$$\begin{bmatrix} I_{SSA}(t) \\ Q_{SSA}(t) \end{bmatrix} = H_{SSA}(s) \left\{ \begin{bmatrix} u_I(t) \\ u_Q(t) \end{bmatrix} + \begin{bmatrix} d_I(t) \\ d_Q(t) \end{bmatrix} \right\} \quad (7)$$

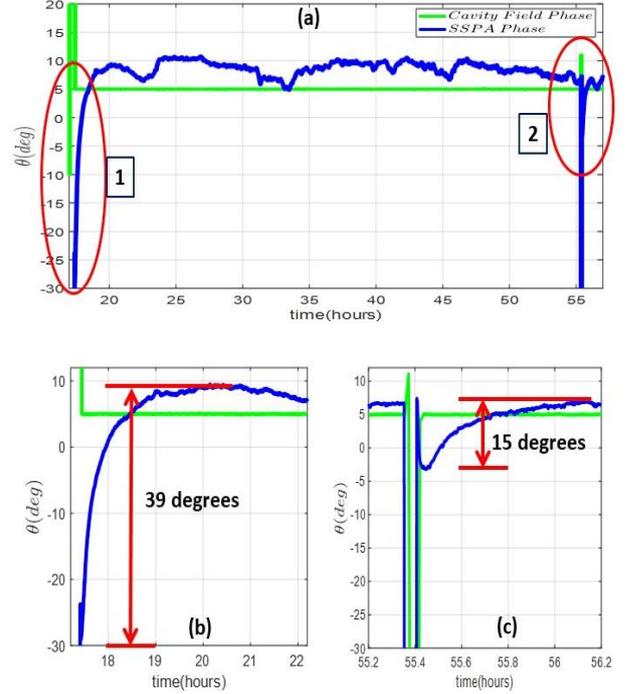

**Figure 4**. Phase drift observation of the SSPA via the DOB. Data was captured at the middle of 1000 usec long RF pulse (a) Longterm (42 hours) Phase drift. (b) (Zoom of box1 in figure 4a) SSPA Phase drift monitored where the RF was recovered from a few hours Turn Off. SSPA was fully cool down. (c) (Zoom of box2 in figure 4a) SSPA Phase drift monitored where the RF was recovered from 10 minute RF trip.

where $H_{SSA}(s)$ is the uncertain model of the SSPA and the phase variation in the SSPA is represented as the input disturbances $d_I(t)$, $d_Q(t)$ (refer to figure 1). The nominal model $H_n(s)$ of the SSPA can be expressed as a gain, a rotation matrix and a first order lowpass filter with the high cutoff frequency,

$$H_n(s) = \frac{h_{SSA}}{\tau_{SSA}s+1} \begin{bmatrix} cos(\theta_{SSA}) & -sin(\theta_{SSA}) \\ sin(\theta_{SSA}) & cos(\theta_{SSA}) \end{bmatrix} \quad (8)$$

where $h_{SSA}$ is the steady state gain of the SSPA, $\tau_{SSA}$ is the time constant of the SSPA, and $\theta_{SSA}$ is the rotation angle inside the SSPA. Applying the DOB [3] to the system (7), the estimates of the disturbances $d_I(t)$, $d_Q(t)$ are obtained as

$$\begin{bmatrix} \hat{d}_I(t) \\ \hat{d}_Q(t) \end{bmatrix} = Q_{ob}(s) H_n^{-1}(s) \begin{bmatrix} y_I(t) \\ y_Q(t) \end{bmatrix} - Q_{ob}(s) \begin{bmatrix} u_I(t) \\ u_Q(t) \end{bmatrix} \quad (9)$$

where $Q_{ob}(s)$ is the so called Q-filter of the DOB.

The nominal model $H_n(s)$ has the dynamics of a first order lowpass filter with the high cutoff frequency, which means that the implementation of the inverse of the nominal model, $H_n^{-1}$ on the LLRF DSP is impossible. Then, instead of $H_n^{-1}$, $Q_{ob}H_n^{-1}$ is implemented with the design of the DOB filter $Q_{ob}(s)$. Here, $Q_{ob}(s)$ is designed as a two-pole lowpass filter. The first pole is for the cancellation of the zero of $H_n^{-1}$. The second pole is for the first order lowpass filter characteristics of $Q_{ob}H_n^{-1}$ so that the DOB can estimate the low frequency disturbance, $d_I(t)$, $d_Q(t)$ and reject high frequency sensor/detector noise. In the case of our application of DOBC to suppress the phase variation of the SSPA, the cutoff frequency $Q_{ob}$ (s) is set at the same frequency of the 3dB bandwidth of the nominal cavity, 2kHz.

Figure 5 shows the DOBC Performance results while 625 usec long H+IP Beam was operated for the Medical Isotope Production. The beam is on at 375 usec. Figure 5a shows cavity field amplitude error and figure 5b shows cavity field phase error. Figure 5a, 5b show that the DOBC improves the amplitude and phase performances discernably at the beam loading transient. Figure 5a, 5b also show that the beam pulse length can be elongated from 625 usec to 750 usec satisfying the error requirements of LANSCE LINAC ($\pm 1.0$ % amplitude error and $\pm 1.0$ degrees phase error). Figure 5c shows the effectual SSPA phase variation, $\theta_d - \hat{\theta}_d$, which shows that at 200 usec, the DOBC improves the effectual phase variation in the SSPA at 5.7 degrees.

## 5. SUMMARY

In this paper, the disturbance observer controller that compensates for the phase variation at the solid state power amplifier of the low energy LANSCE DTL cavity was addressed. The controller stabilized the solid state power amplifier against the 40 degree phase perturbation occurring at the a few tens minute RF turn-on transient, which was the main concern of the operation of the solid state power amplifier, and provided large phase margin to the PI feedback control system. The disturbance observer controller was implemented on the existing DLLRF system of a FPGA without an extra hardware, which is a big benefit of the DLLRF system.

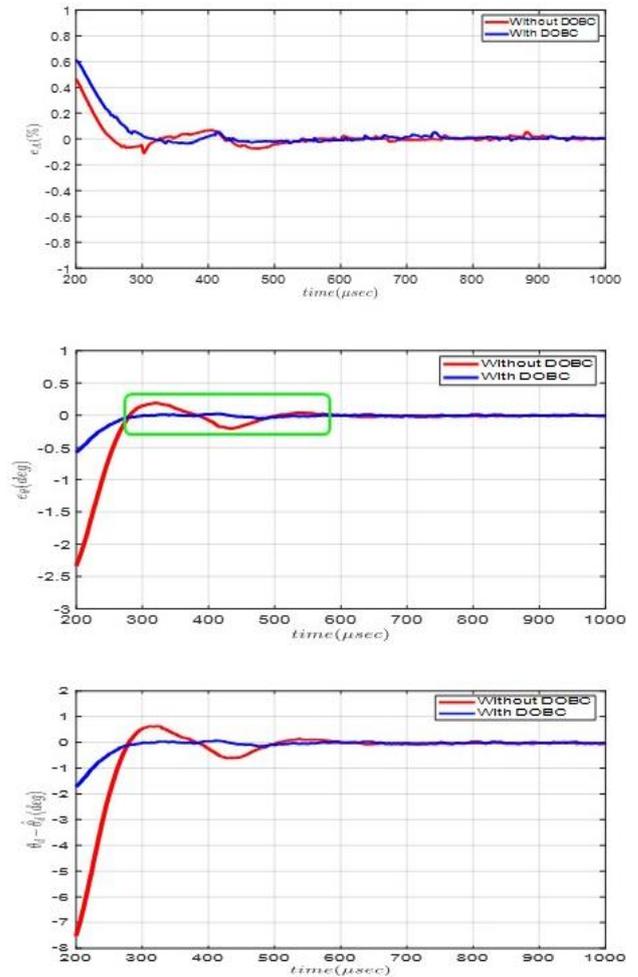

**Figure 5**. DOBC Performance Results of 625 usec long H+IP Beam operation for the Medical Isotope Production.